\RequirePackage{lineno}
\documentclass[twocolumn]{emulateapj}
\usepackage{amsmath}
\usepackage{amssymb}
\usepackage{hyperref}
\usepackage{comment}
\usepackage{color}
\usepackage{natbib}
\usepackage{verbatim}
\usepackage{graphicx}
\usepackage{soul}
\usepackage{rotating}
\usepackage{xspace}
\usepackage{upgreek}

\usepackage{aas_macros}

\newcommand{\lsi}{\,\raisebox{-0cm}{$\stackrel{\textstyle<}
{\textstyle\sim}$}\,}
\newcommand{\gsi}{\,\raisebox{-0.13cm}{$\stackrel{\textstyle> 
}
{\textstyle\sim}$}\,}

\shorttitle{Extended gamma-ray emission from Andromeda --  millisecond pulsar interpretation}
\shortauthors{C. Eckner {\it et al.}}

\begin{document}

\title{Millisecond pulsar origin of the Galactic center excess and extended gamma-ray emission from Andromeda -- a closer look }

\author{
Christopher Eckner\altaffilmark{1,$\star$}, Xian Hou\altaffilmark{2,3,4,$\ast$}, Pasquale D. Serpico\altaffilmark{5,$\diamond$}, Miles Winter\altaffilmark{6},
Gabrijela Zaharijas\altaffilmark{1,7,$\dagger$},
Pierrick Martin\altaffilmark{8}, Mattia di Mauro\altaffilmark{9}, Nestor Mirabal\altaffilmark{10, 11}, Jovana Petrovic\altaffilmark{12}, Tijana Prodanovic\altaffilmark{12}, Justin Vandenbroucke\altaffilmark{5}
}
\altaffiltext{$\star$}{\href{christopher.eckner@ung.si}{christopher.eckner@ung.si}}
\altaffiltext{$\ast$}{\href{xianhou.astro@gmail.com}{xianhou.astro@gmail.com}}
\altaffiltext{$\diamond$}{\href{serpico@lapth.cnrs.fr}{serpico@lapth.cnrs.fr}}
\altaffiltext{$\dagger$}{\href{gabrijela.zaharijas@ung.si}{gabrijela.zaharijas@ung.si}}

\affil{$^1$ Laboratory for Astroparticle Physics, University of Nova Gorica, Vipavska 13, SI-5000 Nova Gorica, Slovenia}
\affil{$^2$ Yunnan Observatories, Chinese Academy of Sciences, 396 Yangfangwang, Guandu District, Kunming 650216, P.~R.~China}
\affil{$^3$ Key Laboratory for the Structure and Evolution of Celestial Objects, Chinese Academy of Sciences, 396 Yangfangwang, Guandu District, Kunming 650216, P. ~R.~China}
\affil{$^4$ Center for Astronomical Mega-Science, Chinese Academy of Sciences, 20A Datun Road, Chaoyang District, Beijing 100012, P.~R.~China}
\affil{$^5$ LAPTh, Universit\'e Savoie Mont Blanc \& CNRS, 74941 Annecy Cedex, France}
\affil{$^6$ University of Wisconsin, Madison, 500 Lincoln Drive, Madison, WI, 53706, USA}
\affil{$^7$ Istituto Nazionale di Fisica Nucleare - Sezione Trieste, Padriciano 99, I-34149 Trieste, Italy}
\affil{$^8$ Institut de Recherche en Astrophysique et Plan\'etologie, CNRS-INSU, Universit\'e Paul Sabatier, 9 avenue Colonel Roche, BP 44346, 31028 Toulouse Cedex 4, France}
\affil{$^9$ W.~W.~Hansen Experimental Physics Laboratory, Kavli Institute for Particle Astrophysics and Cosmology, Department of Physics and SLAC National Accelerator Laboratory, Stanford University, Stanford, CA 94305, USA}
\affil{$^{10}$ CRESST/CSST/Department of Physics, UMBC, Baltimore, MD 21250, USA}
\affil{$^{11}$ NASA Goddard Space Flight Center, Greenbelt, MD 20771, USA}
\affil{$^{12}$ Department of Physics, Faculty of Sciences, University of Novi Sad, Trg Dositeja Obradovi\'ca 4, 21000 Novi Sad, Serbia}

\date{}

\begin{abstract}
A new measurement of a spatially extended gamma-ray signal from the center of the Andromeda galaxy (M31) has been recently published by the Fermi-LAT collaboration, reporting that the emission broadly resembles the so-called Galactic center excess (GCE) of the Milky Way (MW). Steadily, the weight of the evidence is accumulating on a millisecond pulsar (MSPs) origin for the GCE. These elements prompt us to compare the mentioned observations with what is, perhaps, the simplest model for an MSP population, solely obtained by rescaling of the MSP luminosity function determined in the local MW disk via the respective stellar mass of the systems. Remarkably, we find that without free fitting parameters, this model can account for both the energetics and the morphology of the GCE within uncertainties. For M31, the estimated luminosity due to primordial MSPs is expected to contribute only about a quarter of the detected emission, although a stronger contribution cannot be excluded given the large uncertainties. If correct, the model predicts that the M31 disk emission due to MSPs is not far below the present upper bound. We also discuss additional refinements of this simple model. Using the correlation between globular cluster gamma-ray luminosity and stellar encounter rate, we gauge the dynamical MSP formation in the bulge. This component is expected to contribute to the GCE only at a level $\lesssim 5\%$, but it could affect the signal's morphology. We also comment on limitations of our model as well as on future perspectives for improved diagnostics.
\end{abstract}


\maketitle

\section{Introduction}  \label{sec:intro}

The high energy gamma-ray emission from the MW galaxy is dominated by diffuse interstellar emission, often dubbed Galactic Diffuse Emission (GDE). In the last decade, the GDE has been exquisitely mapped by the Fermi LAT (\cite{2009ApJ...697.1071A}) and AGILE (\cite{Pittori:2009es}) satellites, allowing detailed studies of properties of the galactic cosmic ray population and of the interstellar medium (ISM) (see e.g. \cite{Ackermann:2012pya,Trotta:2010mx,Evoli:2012ha}). Such analyses also revealed particularities of the gamma-ray emission from the inner region of our Galaxy. Notably, the Fermi Bubbles (FB), a large hour-glass structure extending up to 8 kpc away from the plane, have also emerged from the LAT data (\cite{Su:2010qj,Fermi-LAT:2014sfa}) after subtraction of the GDE.

Closer to the plane and within 2 kpc from the Galactic Center (GC), a spectrally distinct component (dubbed the 'Galactic Center excess' (GCE)), was identified by a number of authors, and  its existence and basic properties eventually established over a series of works (\cite{Hooper:2013rwa,Calore:2014xka,TheFermi-LAT:2015kwa,Daylan:2014rsa,TheFermi-LAT:2017vmf}).
While the origin of the GCE signal (just like the one of the FB) is still under debate, possible interpretations include emission from unresolved population of millisecond pulsars (MSPs) (\cite{Abazajian:2012pn,Yuan:2014yda,Brandt:2015ula}), a past transient event which injected high energy particles in the center of the Galaxy (\cite{Petrovic:2014uda,Carlson:2014cwa}), additional steady state sources in the central molecular zone (\cite{Carlson:2015ona,Gaggero:2015nsa}) or annihilation of dark matter particles (\cite{Calore:2014xka,Daylan:2014rsa,Huang:2015rlu, TheFermi-LAT:2017vmf}).  
An interpretation in terms of the MSP emission was originally motivated by the spectral similarity of their gamma-ray emission with the GCE and, to some extent, by the fact that MSPs could migrate over kpc distances during their lifetime, roughly compatible with the size of the excess. This interpretation recently gained support due to studies favouring an unresolved point source origin for the GCE rather than a genuinely diffuse emission (\cite{Bartels:2015aea,Lee:2015fea}).  Even within the MSP scenario, a still open question is whether the properties of the MSP bulge and disk populations are the same, for instance in terms of the MSP luminosity function or astrophysical origin. 

In addition to the MW, seven external star-forming galaxies have been detected in gamma-rays with the Fermi LAT, allowing for studies of a correlation between the global galactic properties and overall gamma-ray emission (\cite{Ackermann:2012vca}). M31 takes up a special place among them, as it is the only other large spiral with a prominent bulge and it is close enough so that the disk and bulge could be resolved as two separate components (in infrared, the disk spans $4^\circ$ on the sky).
M31 was first detected in gamma-rays with a $5.3\, \sigma$ significance in the two year Fermi LAT data, with marginal detection of a spatial extension (at $1.8\, \sigma $) (\cite{Fermi-LAT:2010kib}). The most recent analysis, using 88 months of Pass 8 SOURCE class events enabled a more precise measurement (\cite{Ackermann:2017nya}, we will refer to this work as 'PaperI' in what follows).

The new measurement revealed a few key points: (i) the emission was found to be slightly extended, with the significance of extension at the $4\,\sigma$ level; (ii) the morphology of the signal is well described by a uniform disk of a radius $0.38\pm0.05^{\circ}$ or a Gaussian distribution with a width of $0.23\pm0.08^{\circ}$ and the emission does not correlate with regions rich in gas (as traced by an atomic gas column density NH map) or current star-formation activity (traced by Herschel/PACS map at 160 $\upmu$m); (iii) the measured spectrum is consistent with a simple power law but also with a power-law spectrum with an exponential cut-off in the GeV range, the latter more closely resembling MSP-like spectral templates.
The fact that the gamma-ray emission does not correlate with gas or star formation sites can be explained by the currently low star formation rate in M31 ($0.25M_{\odot}\textrm{yr}^{-1}$) which is about ten times lower than in the MW ($2M_{\odot}\textrm{yr}^{-1}$)(\cite{Ford:2013hva}). As a consequence, the low level of the disk emission leaves us with an intriguing possibility to study the inner-galaxy emission of a 'MW-like' galaxy in an unobstructed way.  

In this paper we explore the MSP interpretation of the MW and M31 inner galaxy emission using what is probably the simplest model for the expected MSP luminosity:
As basic hypothesis, we assume that the MSP content essentially correlates with the stellar mass of the object under study, relying on the measured properties of the population of gamma-ray MSPs detected in the MW disk to extrapolate. We anticipate that, remarkably, such a simple model is already capable to account for the energetics (and, to a large extent, morphology) of the GCE, within current uncertainties.
Then, we  apply  the same model to M31, in order to test the possibility of a unique origin of both signals. 
We further discuss the impact of refinements to this model, notably the possibility of an additional MSP population formed dynamically via encounters, which {\it would not} simply  correlate with the stellar mass of the object under study, and calibrated to the observed correlation between the gamma-ray luminosity of globular clusters and the stellar encounter rates in those systems. Other alterations in the scaling between disk and bulge luminosity as well as alternatives on the origin of the MSP are also briefly addressed. 

This article is structured as follows: Basic notions on MSPs needed for our work are discussed in Sec.~\ref{sec:model}. In Sec.~\ref{sec:density} we introduce the astrophysical inputs for the model, notably the stellar density profiles for both the MW and M31, with particular attention to their bulges (and substructures). In Sec.~\ref{sec:MWGCEfit} (respectively, Sec.~\ref{sec:M31_emission}) we characterize the observed MW (respectively, M31) emission, while Sec.~\ref{sec:ModelMW} (respectively, Sec.~\ref{sec:M31-toy-model}) is devoted to the comparison with the model prediction.  Refinements to the simple model, some caveats, a few additional effects and model alternatives are discussed in Sec.~\ref{sec:higher-order-contribution}. In Sec.~\ref{sec:summary} we summarize and conclude. The appendix is dedicated to some considerations on the (still limited) spectral information on the M31 signal, on how it compares to the expectations, and possible tails at high energy.

\section{Basics on millisecond pulsars}\label{sec:model}

MSPs are rapidly spinning  (and slowly evolving) pulsars, often found in binary systems, typically interpreted as old pulsars whose rotational frequency has been boosted through the mass transfer from their companion star. MSPs are found both in the near ISM environment and in old stellar systems, such as globular clusters or galaxy bulges. 
Although it has been proposed that MSPs might have been deposited in galaxy bulges through the infall and tidal disruption of globular clusters (\cite{Gnedin:2013cda, Brandt:2015ula, Fragione:2017rsp}), at least in the main part of this work 
we focus on MSP formation {\it in situ}. We comment on the above alternative in Sec.~\ref{sec:corrections-lum}.

There are two main contributions to the {\it in situ} (\cite{Mirabal:2013rba}) formation of MSPs. The first channel requires a close binary system with an extreme mass ratio of the two stars, which remains bound after the birth of a pulsar in the supernova explosion of the more massive partner. In a later phase, during the mass transfer from the companion star to the neutron star, the binary becomes visible as a low-mass X-ray binary (LMXB). In this stage, the binary system is accretion-powered and shines in X-rays. Eventually, it fades away in a rotation powered phase resulting in an MSP accompanied by a low-mass white dwarf. This MSP formation mechanism, known as the  {\it primordial} channel, also implies that the numbers of MSPs in a given system is proportional to the total stellar mass, at least if the fraction of binaries is an unbiased tracer of stellar mass. 
In galaxy disks, for example, where the primordial formation channel is expected to be dominant, the number of LMXBs scales linearly with stellar mass of the host galaxy (\cite{2004MNRAS.349..146G}), landing an observational support to this scaling relation. Some of the authors of this work already used the scaling with the mass of the MSP population in the MW disk to predict the MSP population in dwarf satellite galaxies of the MW (\cite{Winter:2016wmy}).

But MSPs can also form by neutron star {capture} into a binary system through stellar encounters (\cite{1998MNRAS.301...15D,Hui:2010vt}), a process  dominant in globular clusters, where stellar densities are much higher than in the disk. Interactions leading to the dynamical formation of MSPs depend on bringing two stellar systems (like a neutron star and a binary) close together, and thus depend on the square of the stellar density. At the same time, gravitational focusing is reduced by the stellar velocity dispersion, leading to a dependence of the stellar encounter rate $\Gamma_e$ on cluster properties as $\rho_{\ast} ^2/ \sigma$, where $\rho_{\ast}$ is the stellar density and $\sigma$ is the velocity dispersion (\cite{1975MNRAS.172P..15F,Bahramian:2013ihw}). This so called {\it dynamical} formation channel is supported by globular cluster observations (see (\cite{collaboration:2010bb,2011ApJ...726..100H})).

In what follows, we first focus on the expected luminosity of primordial MSPs in both the MW and M31 bulge, based on the properties of MSP inferred from the local (MW disk) measurements. In Sec.~\ref{sec:higher-order-contribution}, we estimate and comment upon the contribution from dynamically formed MSPs.

\begin{table*}[!ht]
\caption{}{Summary of sizes and stellar masses of MW and M31
subcomponents. The estimate of the spatial extension of M31's stellar disk is taken from (\cite{2011ApJ...739...20C}) whereas the MW's stellar disk size is reported in \cite{Rix:2013bi}. References to the listed stellar masses are given in the text. \label{tbl:sizes_subcomponents}}
\begin{center}
\begin{tabular}{l c c c}
\tableline\tableline
\\
& Component & spatial extension & stellar mass $\left[M_{\odot}\right]$\\ 
\\
\tableline
\tableline
 MW $\vphantom{\intop_x^x}$ ($r_{\odot} = 8.5\;\textrm{kpc}$)\\
\tableline
\tableline
& disk $\vphantom{\intop_x^x}$ & 30 kpc & $6\times10^{10}$\\
& bulge &  2 kpc & $\left(9.1\pm0.7\right)\times10^{9}$  \\
& NSC &  10 pc &  $\left(2.5\pm0.4\right)\times10^{7}$\\
\tableline
\tableline
 M31 $\vphantom{\intop_x^x}$ (distance: $d = 785\;\textrm{kpc}$)\\
\tableline
\tableline
& disk $\vphantom{\intop_x^x}$& $15-56$ kpc & $7-10\times10^{10}$\\
& bulge &  2 kpc  & $\left(4.0\pm1.0\right)\times10^{10}$ \\
& NSC & 10 pc  &  $3.5\times10^{7}$\\
\tableline \tableline
\\
\\
\\
\end{tabular}
\end{center}
\end{table*}
\section{Stellar densities: Inputs} \label{sec:density}

For the stellar density in the MW bulge we adopt (\cite{2009arXiv0903.0946V})
\begin{equation}
\rho _{\textrm{MW,b}} (a)= \rho_{0,\textrm{MW}} \frac{e^{-a^2/a_m^2}}{\left(1 + a/a_0 \right)^{1.8}}\,,\label{rhobulgeMW}
\end{equation}
where $a$ is the distance from the GC, the parameters $a_m=1.9$ kpc and $a_0=100$ pc are taken from (\cite{1997MNRAS.287L...5B}), while $\rho_0$ is obtained via the normalization to the total mass of the bulge, which we take to be $M_{\textrm{MW,b}}=\left(0.91\pm0.07\right)\times10^{10}M_\odot$ (\cite{Licquia:2014rsa}). 
We use (\cite{Licquia:2014rsa}) as an input since it provides numbers for both the bulge and the bulge+disk stellar mass in a single publication. Given that the ratio between the two values is more important for our analysis than the absolute value of the bulge mass, we make sure that both masses are determined at least self-consistently. However, note  that the value for the bulge mass reported varies almost by a factor of two when compared to (\cite{Portail:2016vei}). 

Also, in Eq.~\ref{rhobulgeMW} we assume that the stellar distribution is spherical, although the MW bulge is rather elliptical, and includes a bar. Since  in what follows we compare our model predictions with the observed latitude profile of the GCE, calculated in cases in which spherical symmetry is imposed, this approximation is self-consistent. Additionally, the migration of the MSPs (see Sec.~\ref{sec:ModelMW}) would go in the direction of smoothing the actual profile, making it closer to spherical. Under the definition of the stellar density above, the radius of the volume which contains half of the mass is 1.2 kpc. In order to account for most of the stellar mass we take the size of the MW bulge to be 2 kpc in what follows,  corresponding to approximately 75\% of $M_{\textrm{MW,b}}$.

In order to estimate the contribution of dynamically formed MSPs later on (Sec.~\ref{sec:higher-order-contribution}), it is important to account for the enhanced densities in the central $\mathcal{O}\!\left(10\right)$ pc, the so called Nuclear Star Cluster (NSC). The MW's NSC was found to be point-symmetric about its center, Sgr $\rm{A}^{\ast}$, while exhibiting a flattening along the Galactic plane with minor-to-major projected axis ratio $q = 0.71\pm0.02$. Extending to approximately 10 pc, the NSC encompasses a total mass of $\left(2.5\pm0.4\right)\times10^7\,M_{\odot}$ (\cite{Schodel:2014gna}).
 
The stellar density in M31 is adopted from (\cite{2012A&A...546A...4T})
\begin{align}
\rho_{\textrm{M31,b}} (a) = \rho_{0,\textrm{M31}} ~e^{-d_N \left[(a/a_c)^{1/N} -1 \right]}\,,
\end{align}

with $d_N=7.769$, $a_c=1.155$ kpc and $N=2.7$. {Here the distance $a_c$  marks the volume which contains half of the mass of the bulge. We obtain $\rho_{0,\textrm{M31}}$ by normalizing to the total mass of the M31 bulge, taken to be $\left(4.0\pm1.0\right)\times 10^{10}$ M$_{\odot}$, as in (\cite{1989AJ.....97.1614K}). We perform our integrations up to 2 kpc in what follows, noting that M31 and MW stellar bulges have comparable volumes.

A modelling of the NSC of M31 is specially challenging as it is known to exhibit a complex morphology. The inner 1.8 pc region of the NSC in M31 features a bimodal shape (Lauer et al. 1993). It can be interpreted as a projection of a central eccentric disc (Peiris \& Tremaine 2003) and can be explained by the fact that the radius of gravitational influence of the Super Massive Black Hole in M31---contrarily to the MW case---extends over the half  of the volume of NSC, affecting the shape of the NSC in a non trivial way (\cite{2016MNRAS.457.2122G}). We use a simple parametrization of the density of the NSC from (\cite{2012A&A...546A...4T}), but caution the reader that uncertainties are significant.

A summary of sizes and stellar masses of all studied components of the MW and M31 is provided in Tab.~\ref{tbl:sizes_subcomponents}.

\section{MW Inner Galaxy emission} \label{sec:MWGCEfit}

\subsection{Overall luminosity}

To calculate the luminosity of the GCE we use the latest analysis of the region with Fermi LAT's 6.5 year Pass 8 events of the ULTRACLEANVETO class, as derived in (\cite{TheFermi-LAT:2017vmf}).
This work presents a detailed exploration of systematic uncertainties on the derivation of the properties of the excess, including: possible additional sources of CRs near the GC, tests of the low-latitude morphology of the FB and of the GCE, uncertainties of the predictions of the inverse Compton emission and of the gas column densities. 
By taking this wide range of uncertainties into account, \cite{TheFermi-LAT:2017vmf} derive the total flux of the GCE emission within the inner $10^\circ$ from the GC. There, the GCE flux is derived by masking all resolved point sources in the region, which account for $20\%$ of the ROI. Assuming a 8.5 kpc distance to the GC  and taking into account the masked region, we arrive to the luminosity of the GCE in this region above 100 MeV of:

\begin{equation} \label{eq:lum1gce}
L_{{\rm GCE}} ^{10^\circ}= 1.3^{+0.9} _{-1.1}\times10^{37}\text{erg s}^{-1}\,.
\end{equation}

If we focus on the special case for which a spatial template with the MSP-like spectrum was used in the analysis, the total luminosity becomes $(0.2\pm 0.1)\times10^{37}$ {erg s}$^{-1}$, where the quoted error is statistical. We caution however, that the luminosity of this component was obtained in a simultaneous fit with the template of FB, which was allowed to vary in the intensity in the inner 10$^\circ$ region, and could have potentially absorbed part of our signal. For that reason, this luminosity could be regarded as the lower limit to the MSP emission.

We assume that this luminosity originates in a population of individually unresolved MSPs, as supported by increasing amount of evidence (\cite{Bartels:2015aea,Lee:2015fea}). 

We also need to make sure that we capture the majority of the MSP emission associated with the bulge. Beyond $10^\circ$ from the GC, the GCE emission becomes comparable to the emission from the FBs so the total extent is hard to determine robustly. As shown in (\cite{Brandt:2015ula}), the emission from the region $10^\circ - 20^\circ$ (a ROI of $20^{\circ}$ corresponds to a region with radius $\sim3.1\;\textrm{kpc}$ around the GC) is expected to add about $30\%$ to the emission coming from the inner region, so that 

\begin{equation} \label{eq:LobsGCEtotal}
L_{{\rm MSP}} ^{20^{\circ}}\simeq 1.3\,L_{{\rm GCE}} ^{10^\circ}=  1.7^{+1.2} _{-1.4} \times10^{37}\text{erg s}^{-1}\,.
\end{equation}
Based on the new catalog of point sources in the inner galaxy derived in (\cite{Fermi-LAT:2017yoi})\footnote{Note that the cited point source catalog was derived in v1 of this paper which still remains correct whereas the originally derived pulsar luminosity function was corrected in v2 of (\cite{Fermi-LAT:2017yoi}). See also (\cite{Bartels:2017xba}).}, we estimate that the contribution of individually resolved MSPs to this luminosity does not exceed a 20\% effect, well below the estimated error.

\subsection{Morphology}

In (\cite{TheFermi-LAT:2017vmf}) the best fit to the data was found to be given by a template with the emission decreasing like $r^{-2}$ (corresponding to the $r^{-1}$ NFW dark matter density profile, for annihilating DM), while earlier analysis (\cite{Daylan:2014rsa,Calore:2014nla}) were finding a steeper slope of $r^{-2.4}$ (dubbed 'generalised' NFW, or gNFW).  

In (\cite{TheFermi-LAT:2017vmf}) the authors also derived a spatial distribution of the emission which exhibits the MSP-like spectrum. We note that this profile is derived in an analysis in which the template for the hard FB-like spectral emission was fitted to the data simultaneously, which could influence the derived morphology within 20$^\circ$, where the FB template was found to be significantly brighter than at higher latitudes.
We report the measurements of the radial profile of GCE mentioned above, as well as other determinations from previous literature in Fig.~\ref{fig1}.

\begin{figure*}
\centering
\includegraphics[width=.65\textwidth]{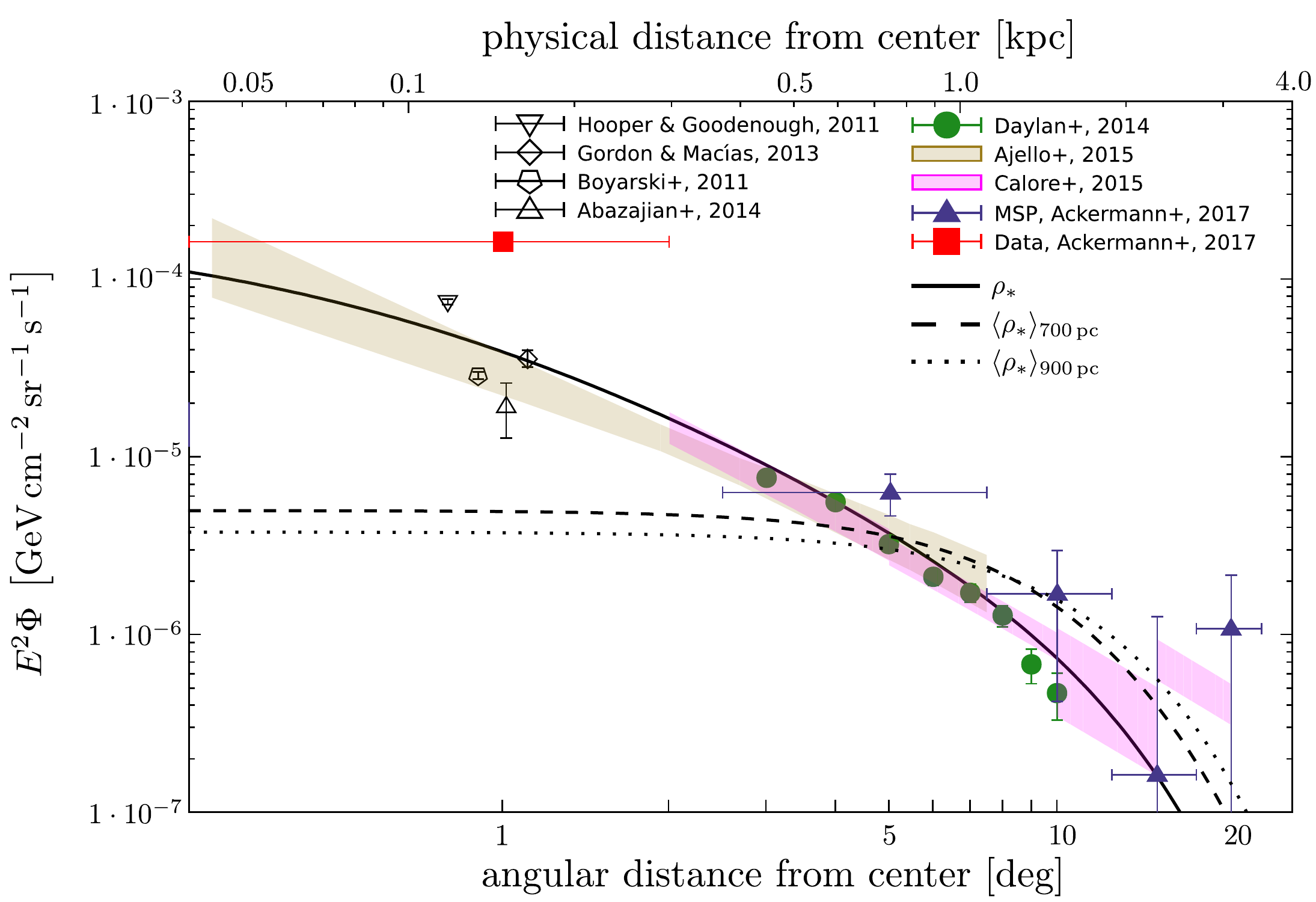}
\caption{Radial profile of the GCE emission, as derived in a series of works, shown together with the predicted flux from a population of primordially formed MSPs in the Galactic bulge for three different smoothing kernels $d =$ 0 pc (solid), 700 pc (dashed) and 900 pc (dotted) applied to the stellar density in the MW bulge. The uncertainty on the normalization of the predictions is given in Eq. \ref{eq:Lmwprim}, with the actual curves shown only reporting 
the best fit agreement to \cite{Calore:2014xka} within the allowed predicted range to avoid cluttering. }\label{fig1}
\end{figure*}

\section{Prediction for the MW} \label{sec:ModelMW}
\subsection{Overall luminosity}
To calculate the expected MSP emission in the MW bulge by considering the primordial formation channel, we start by making the simple hypothesis that the total number of MSPs in a given object is proportional to the (stellar) mass of the system. A number of derivations of the luminosity functions  of the MSP in the MW disk has been reported in the literature, where spatial distributions roughly (but not exactly) matching this hypotheses have been adopted. Here we follow and adopt the results of (\cite{Winter:2016wmy}), estimating the MSP luminosity in the MW bulge via the simple rescaling:
\begin{equation} \label{eq:prim}
 L_{\rm b}^{\rm MW}=   \frac{M_{\rm b}^{\rm MW}}{M_{\ast}^{\rm MW}} \int_{L_{\rm min}}^{L_{\rm max}} L_\gamma \left( \frac{dN}{dL_\gamma} \right)_{\rm MW} dL_\gamma \,,
\end{equation}

where $M_{\rm b}^{\rm MW}$ and $L_{\rm b}^{\rm MW} $ are the stellar mass and the predicted cumulative luminosity of the bulge of the MW, while $M_{\ast}^{\rm MW}$ is the stellar mass of the MW disk.  The luminosity function of the MW, $ \left( \frac{dN}{dL_\gamma} \right)_{\rm MW}$, was obtained by considering 66 observed field MSPs\footnote{\url{https://confluence.slac.stanford.edu/display/GLAMCOG/Public+List+of+LAT-Detected+Gamma-Ray+Pulsars}} and by taking into account the incompleteness of MSP detection in the Galaxy by the Fermi LAT. The main contribution to the systematic uncertainty on the luminosity function comes from the detection incompleteness.  Similar analyses where performed in \cite{Yuan:2014rca, hoo15} and more recently in (\cite{Fermi-LAT:2017yoi, Ploeg:2017vai, Bartels2017}), reaching different results depending on the assumptions, although consistent with each other within the (large) uncertainties.

We assume the stellar mass of MW's disk to be $M_{*}^{\rm MW}=6\times10^{10}M_\odot$ (\cite{Licquia:2014rsa}), and account for the uncertainty in the bulge mass and luminosity function below. 
Assuming $M_{\rm b}^\textrm{MW}=0.91\pm0.07\times10^{10}M_\odot$ (\cite{Licquia:2014rsa}), we obtain:
\begin{equation} \label{eq:Lmwprim}
 L_{\rm b}^{\rm MW}= 1.7^{+2.6}_{-1.0}\times10^{37}\text{erg s}^{-1}\,.
\end{equation}
By comparing with Eq.~\ref{eq:LobsGCEtotal} we see that, at least bolometrically, the GCE luminosity can be explained by the MSPs emission via a simple rescaling.\\
The MSP luminosity function of the MW is however not well-constrained, also depending to some extent on the assumptions made to derive it (most notably on the model of the spatial distribution of MSPs in the Galaxy, and on the uncertainty on the MSP distance measure). To reflect the  large uncertainties among different analysis in the literature,  in Tab.~\ref{tbl:MW-bulge-MSP-lum-Ploeg} we report the MW disk luminosities derived in \cite{Ploeg:2017vai, Bartels2017}, obtained from other MSP luminosity functions, comparing them with our baseline estimate. We also  rescale them to the mass of the MW bulge we adopt in this study, showing roughly what prediction those models would have yielded assuming a linear scaling. The same results are show in graphical form in Fig.~\ref{fignew}, where the gray band represent our baseline model for the disk luminosity and its rescaled value in the bulge (bottom and top panel, respectively).

\begin{table*}
\caption{}{Summary of the predicted MSP luminosities of the MW bulge $L_{\textrm{b}}^{\textrm{MW}}$ for different MSP luminosity functions taken from \cite{Winter:2016wmy,Ploeg:2017vai, Bartels2017}. In order to obtain the predictions for \cite{Ploeg:2017vai,Bartels2017}, we rescaled the given luminosity of the MW disk ($L_{\ast}^{\textrm{MW}}$) to the mass of the MW bulge (for adopted mass values see Tab.~\ref{tbl:sizes_subcomponents})} \label{tbl:MW-bulge-MSP-lum-Ploeg}
\begin{center}
\begin{tabular}{l c c}
\tableline\tableline
\\
MSP luminositiy function & $L_{\ast}^{\textrm{MW}}\;\left[10^{37}\,\textrm{erg}\,\textrm{s}^{-1}\right]$ & $L_{\textrm{b}}^{\textrm{MW}}\;\left[10^{37}\,\textrm{erg}\,\textrm{s}^{-1}\right]$ \\ 
\\
\tableline
\tableline
\\
\cite{Winter:2016wmy} &  $11^{+17}_{-6}$ & $1.7^{+2.6}_{-1.0}$ \\
\\
\cite{Ploeg:2017vai} (spherical MW bulge) &   $5.0^{+5.0}_{-2.5}$ & $0.76^{+0.76}_{-0.38}$ \\
\\
\cite{Ploeg:2017vai} (X-shaped MW bulge)&   $7.7^{+12.3}_{-4.7}$ & $1.2^{+1.9}_{-0.7}$ \\
\\
\cite{Bartels2017} & $0.89^{+12.93}_{-0.38}$ & $0.16^{+2.35}_{-0.07}$\\
\\
\tableline \tableline
\\
\\
\\
\end{tabular}
\end{center}
\end{table*}

\begin{figure*}[!ht]
\centering
\includegraphics[width=.5\textwidth]{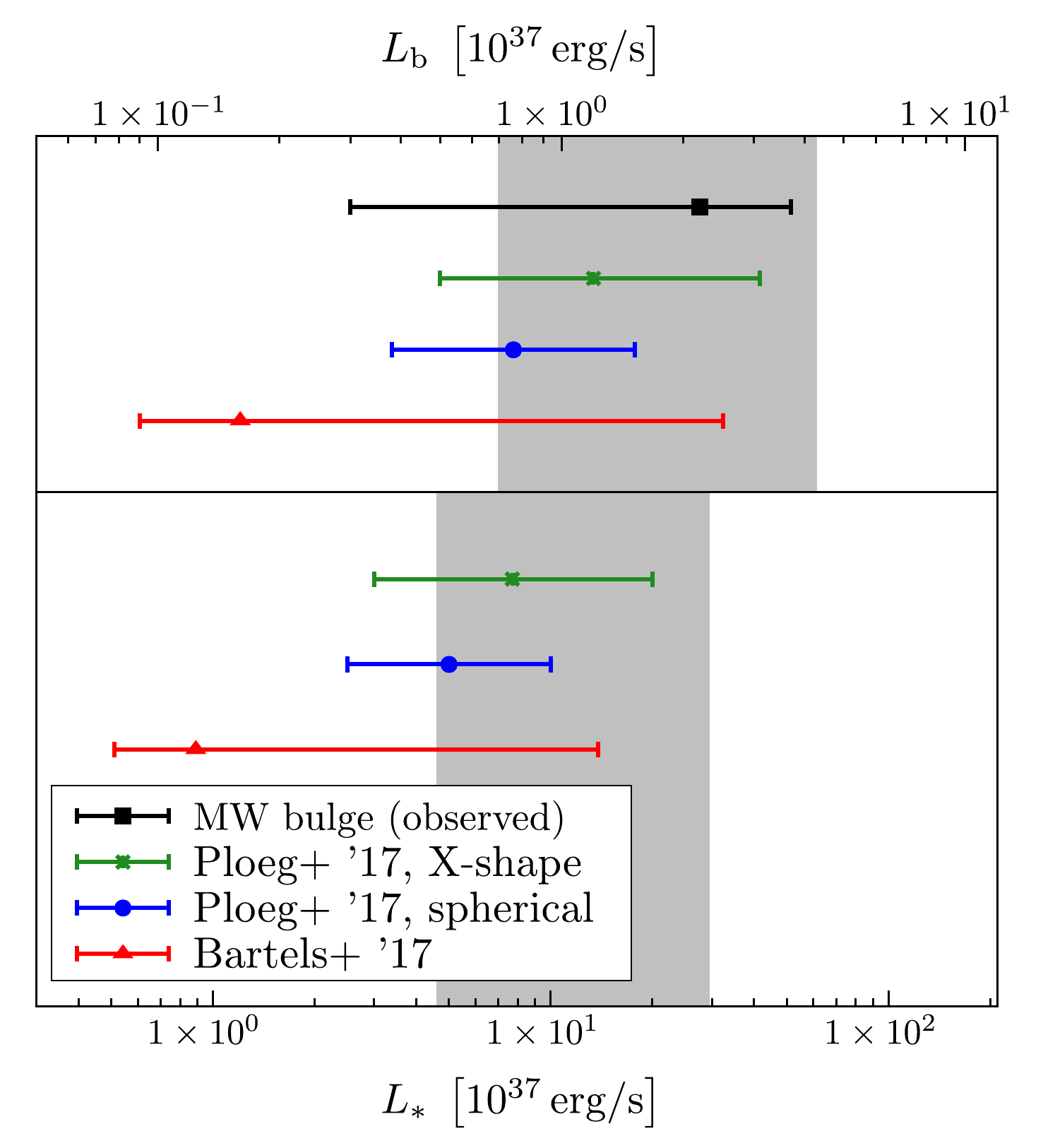}
\caption{Values of the luminosity of the MW's disk (bottom panel) component deduced in (\cite{Ploeg:2017vai}), 
for two geometry assumptions for the bulge (private communication from the authors), and \cite{Bartels2017} compared with the naive prediction for
the MW bulge luminosity obtained via simple rescaling with the mass. The gray band represent our reference model from~\cite{Winter:2016wmy} for the disk luminosity and its rescaled value in the bulge (bottom and top panel, respectively). Our estimate for the MW bulge luminosity is also reported (top point in the top panel).}\label{fignew}
\vspace{0.65cm}
\end{figure*}

\subsection{Morphology}
In order to discuss the morphology, we need to introduce some additional notions. 
Pulsars are known to receive kick-velocities at births, associated to supernova explosions, leading to  ${\cal O}$(1) kpc displacements during their lifetimes (see e.g. (\cite{Lorimer:2005bw,Abazajian:2012pn})). Although measurements concerning isolated pulsars are consistent with a Maxwellian distribution with sizable dispersion $\sigma_{\rm{psrs,~isolated}}=190$ km s$^{-1}$ (\cite{Hansen:1997zw}), the kick velocity distribution of MSPs varies widely in the literature: (\cite{Hooper:2013nhl})  and (\cite{Cordes:1997my}) find $10-50$ km/s , (\cite{2004IAUS..218..139H}) report $85 \pm 13$ km/s, while (\cite{1998MNRAS.295..743L}) suggest a higher velocity range $130\pm30$ km/s. The fact that MSP are found in globular clusters, which are associated to a relatively shallow gravitational potential, suggests that a non-negligible fraction of the population has relatively modest kicks. As reviewed in~\cite{Ivanova:2017ifz}, one possible explanation of the puzzle is that the neutron star population retained in globular clusters may be originated mostly (at $\sim97\%-99\%$!) from electron capture supernovae, associated to progenitors in a narrow range of masses around 8 $M_\odot$, hence to low velocity neutron star kicks. The disk population of neutron stars, on the other hand, would be mostly ($\simeq 90\%$) due to more ``conventional'' core-collapse supernovae. One may also speculate that the situation in the Galactic bulge is intermediate between these two extreme cases.

Anyway, if the kick is insufficient to make the pulsar unbound, it acts in ``broadening'' the pulsar spatial distribution with respect to the progenitor one. In a potential associated to a given mass profile $M(r)$, the virial theorem suggests that the typical size of the new pulsar orbit $d$ is related to the  progenitor one $s$ and the velocity dispersion  $ \langle v^2 \rangle$ as
\begin{equation} \label{eq:virial}
G\left( \frac{M(d)}{d}-\frac{M(s)}{s} \right)= \langle v^2 \rangle\,.
\end{equation}
The quantity $d$ can be thought of as a ``smoothing length'' of the MPS distribution with respect to the underlying stellar density profile.
For MSPs with kicks lower than $\lsi 70$ km s$^{-1}$, the typical smoothing scale for MSPs will be in the range $700-900$ pc and is fairly insensitive to the birth position $s$ (for $s \gsi 300$ pc). For velocity dispersions $\langle v^2 \rangle\gtrsim (70$ km s$^{-1}$)$^2$ a sizable fraction of the MSP can leave the bulge region of interest. Needless to say, a proper calculation of the MSP distribution in the bulge should treat the problem in phase space, but given the preliminary nature of our estimate, we will limit ourselves to show the impact of a $700-900$ pc ``dispersion'' (Gaussian smoothing kernel) due to kicks on the expected signal morphology. A further discussion of the impact of MSP kick-velocities is given in Sec.~\ref{sec:refined_model}.\\

In Fig.~\ref{fig1} we show the radial profile predicted for the cases $d=0, ~700$ pc and $900$ pc. Independently of the smoothing parameters, we see that the profile of the GCE above $\simeq 3^\circ$ is well-reproduced by this simple model and, in the limit of small $d$, the predicted profile matches the observational constraints down to at least the $1^\circ$-scale (note that the top, red-square point is an upper limit to the GCE signal.)  
Despite the large uncertainties that still plague both observations and predictions, it is somehow remarkable that a simple scaling argument, without any major tunable parameter, accounts for both the normalization and geometric profile of the GCE. These two observables add to the original match of the energy spectrum, stimulating further interest in these models. 
In fact, \cite{Macias:2016nev, Bartels2017} confirm the GCE matching of a template tracing the stellar mass. The latter work is based on a refined statistical analysis using the tool \textsc{SkyFACT} to examine  almost 8 years of Fermi-LAT data from the GC region, using a decomposition of the MW bulge into a boxy bulge (\cite{2013MNRAS.434..595C}) and a nuclear bulge (\cite{Launhardt:2002tx}). Although in their study they do not rely on an extrapolation of the disk luminosity, a consensus on the fact that the GCE morphology is consistent with a tracer of the stellar population is clearly emerging. Compared to~\cite{Bartels2017}, in this work we are more concerned about a comparative study of the MW and M31 bulges and disks, as well as  the implications and expectations of different astrophysical  scenarios. Actually, the above discussion indicates that a detailed astrophysical modeling is probably mandatory to make full sense of morphological data, a point not further stressed in~\cite{Bartels2017}. This also justifies the more robust overall luminosity comparison and a ``calorimetric'' perspective we take in our remaining analysis.

\section{M31 emission}\label{sec:M31_emission}

\subsection{Overall luminosity}

As the view on the M31 central region appears unobstructed by the M31 disk, which is significantly fainter in gamma-rays (see Sec.~\ref{sec:M31disk} below), we directly take the luminosity from the M31 central extended source measurement in PaperI and adopt the distance to M31 of 785 kpc reported therein. We will make an assumption (justified {\it a posteriori}) that this emission comes from a population of MSPs associated with the bulge of M31, and we therefore take the emission flux from M31, derived assuming the MSP-like spectrum with a Gaussian spatial template. Hence, we find

\begin{equation}
L_{\rm b, obs}^{\rm M31}= (28. \pm 4.)\times10^{37}\text{erg s}^{-1}\textrm{.} \label{M31obs}
\end{equation}

\subsection{Morphology} \label{profile}

The morphology analysis of M31 is challenging due to the fact that the Point Spread Function (PSF) of the LAT is rapidly changing in the 100 MeV - 10 GeV energy range. In particular, for the P8R2 SOURCE, FRONT+BACK events, used in PaperI, the $68\%$ confinement radius ranges from $1^\circ$ (at 1 GeV) to $0.2^\circ$ (at 10 GeV)\footnote{\url{https://www.slac.stanford.edu/exp/glast/groups/canda/lat\_Performance.htm}}.

In PaperI several spatial templates of the gamma-ray emission were tested and the best fits were found for a uniform disk  with radius $0.38\pm0.05^{\circ}$ ($5.2 \pm 0.7 $ kpc) or a Gaussian distribution with $0.23 \pm 0.08$ deg ($3.1\pm 1.1 $ kpc) extent. A template tracing old stars, which reside dominantly in the bulge (the Spitzer/IRAC maps at 3.6 $\upmu$m), was also tested from which followed that, while it provides a better fit than gas and star formation tracers in M31, it is not favoured compared to a simple point source at the center of M31.

\section{Prediction for M31}\label{sec:M31-toy-model}

We proceed as in Sec.~\ref{sec:ModelMW} and focus on the primordial formation term. Taking $M_{\rm b}^{\rm M31}=4.0\pm1.0\times10^{10}M_\odot$ for the M31 bulge leads to:
\begin{equation} \label{eq:M31primlum}
L_{\rm b}^{\rm M31}= 7.5^{+12.0}_{-5.3}\times10^{37}\text{erg s}^{-1}\,.
\end{equation}

The prediction of MSP luminosity in M31 bulge falls short from the observed value in Eq.~(\ref{M31obs}), though it is consistent at a $2\,\sigma$ level due to the large uncertainties.

\subsection{M31 Disk} \label{sec:M31disk}

It is important to test our model against the `non detection' of the emission from the disk of M31. In the disk, only the primordial formation is relevant and we resort to rescaling of the luminosity function with masses of M31 and MW disks (Eq.~ \ref{eq:prim}). In this way we obtain 

\begin{equation}
L _{*}^{\rm M31}= (1.2-2.7) \times10^{38}\text{erg s}^{-1}\label{eq:lum_M31_fermi}\,.
\end{equation}

In terms of flux, this translates to a value of $(3-4) \times 10^{-12}$ erg/${\rm cm}^2$/s. The upper limit on the total flux from the M31 disk is $4.2 \times 10^{-12}$ erg/${\rm cm}^2$/s (PaperI). While consistent with the expected flux, it is interesting to note that in this simple scenario, a modest improvement over the current bound on the M31 disk emission is expected to lead to a detection. Note, however, that due to the radially decreasing stellar profile, part of the ``extended emission'' detected in M31 may be due to the inner disk stellar MSP population. Although we estimate that this contribution is a sub-dominant fraction of the luminosity reported in  Eq.~\ref{M31obs}, it could certainly contribute sizably to the signal, and should be taken into account in more refined morphological studies. In particular, about $\sim2$ kpc from the Center of M31 the stellar densities of the bulge and disk region are comparable (cf.~\cite{2012A&A...546A...4T}, Fig.~3 therein); we thus expect the largest modification to a simple bulge-like modelling of the signal to be noticeable around that distance.

\subsection{Consistency with M33 upper limit}

Taking the stellar mass of M33 to be $(4.5\pm1.5)~10^9$ M$_\odot$ (\cite{Corbelli:2003sn}), we get a prediction for MSP luminosity (above 100 MeV) of
\begin{equation}
L^{\rm M33} = (8.4^{+14.} _{-6.2})\times10^{36} {\rm erg s}^{-1} \label{eq:lum_M31_prim}
\end{equation}
when integrated over the luminosity interval ($10^{31}-10^{36}$ erg/s). 
Assuming the distance to M33 of 847 kpc (similar to PaperI) we get the total expected flux from M33 of $1.0\times10^{-13}$ erg cm$^{-2}$ s$^{-1}$, which is more than an order of magnitude below the upper limit on the flux from M33 ($< 1.7\times10^{-12}$ erg cm$^{-2}$ s$^{-1}$).

\section{Refinements and additional effects} \label{sec:higher-order-contribution}

Given a population of unresolved, primordially formed MSPs in the MW bulge, we find that their accumulated gamma-ray emission can explain the energetics of the GCE and, to a large extent, also its morphology. Applying the similar rescaling to external galaxies, the model easily accounts for the lack of detection from M33, is also compatible with the upper limit from M31 disk (and, interestingly, it predicts a signal not much below current bound), but can only marginally explain the detected flux from 
 M31 bulge, although it predicts that a sizable fraction of it should be attributed to MSPs. More likely, MSPs do not saturate the measurements, at least in the simplified model discussed above. In the following section, we examine a number of effects that can alter the expectations for both the GCE and M31 signals.

\subsection{Additional $\gamma$-ray sources in M31}

Judging from the complexity of the central region of the MW, it is quite likely that the total luminosity of the extended M31 emission includes further contributions. For instance, there might be a contribution from the central point source of M31. In our own Galaxy, the gamma-ray flux attributed to the GCE is an order of magnitude below the dominating $\pi^0$ component from cosmic ray interactions with interstellar gas, and a factor a few below the inverse Compton component from cosmic ray leptons. It is instead comparable with the flux from resolved point sources at GeV energies, see Fig.~1, right panel in~\cite{TheFermi-LAT:2017vmf}. Assuming that most cosmic rays originate from supernova events, whose rate is dominated by core-collapse events associated to young stars, the $\pi^0$ component is expected to correlate with the recent star-formation activity, which is roughly one order of magnitude lower in M31 than in the MW~(\cite{Ford:2013hva}). Thus, even accounting for the milder non-thermal emission activity in M31 with respect to the MW, it does not appear very surprising that our prediction in Eq. \ref{eq:M31primlum} undershoots the full measured emission by a factor $\lsi 4$, since diffuse emissions in M31 should have a comparable contribution to that of unresolved MSPs. Additional sources,  or morphological differences in the sources relevant in M31 vs. MW (e.g. related to the different spatial distribution of supernova of type I and type II) may also accommodate better the ambiguous and still inconclusive morphological information (Sec.~\ref{profile}).

\subsection{Including dynamical formation} \label{sec:refined_model}

In the central parts of massive galaxies (i.e. inner galaxy bulges), stellar densities can reach values similar to the densities in some globular clusters. In addition, a larger volume with intermediate densities between galactic disks and globular cluster ones also suggests some sizable contribution for a dynamical MSP formation mechanism. Intriguingly, it was observed that the LMXB distribution in M31 follows the $\rho_{\ast}^2$-distribution in the inner region of the M31 bulge, and the $\rho_{\ast}$-distribution in outer regions of the bulge (\cite{Voss:2007hj}). Thus, it seems reasonable to include this contribution to the expected MSP number density $n_{\rm MSP}$ in a formal toy model

 \begin{equation}
n_{\rm MSP}(r) = \langle \rho_{\ast} (r) \rangle/M_{\ast} + \kappa~ \langle \rho_{\ast} (r) ^2 \rangle / \sigma \label{eq:primdyn}\,,
\end{equation}
 
where $\rho_{\ast} $ and $\sigma$ are stellar density and velocity dispersion of a system in question, $M_{\ast}$ refers to the total stellar mass in this system and $\kappa$ is a scaling constant derived from the stellar encounter rate in globular clusters (cf.~Sec.~\ref{sec:globsclustscaling}). The first term on the right-hand side of Eq.~\ref{eq:primdyn} models the primordial formation which was discussed in Sec.~\ref{sec:model} while the second one describes the dynamical formation. Brackets indicate smoothing corrections (possibly different for the two terms) due to the pulsar kicks. In what follows, however, we apply no smoothing to the $\rho_{\ast}^2$ term, since: 
i) the stellar capture is dominated by the low velocity tail of the velocity distribution in the inner regions and ii) there is an observational evidence that the distribution of dynamically formed LMXBs follows a $ \rho_{\ast} ^2$ distribution in M31 (\cite{Voss:2007hj}), a fact that would be spoiled if these corrections were large. 

The $\sigma$ term in Eq.~\ref{eq:primdyn} accounts for the velocity dispersion of neutron stars which are to become captured into a binary system. Stellar rotation and velocity dispersion measurements for the M31 and MW bulges are summarized in (\cite{Widrow:2003yi}) and (\cite{Portail:2016vei}) respectively. The observed velocity dispersions in the two systems seem comparable, being in the $\sim 100- 150$ km s$^{-1}$ velocity range. Given this similarity, we do not take this term into account as it would have a subleading effect on our results.

\begin{figure*}[!ht]
\centering
\includegraphics[width=.65\textwidth]{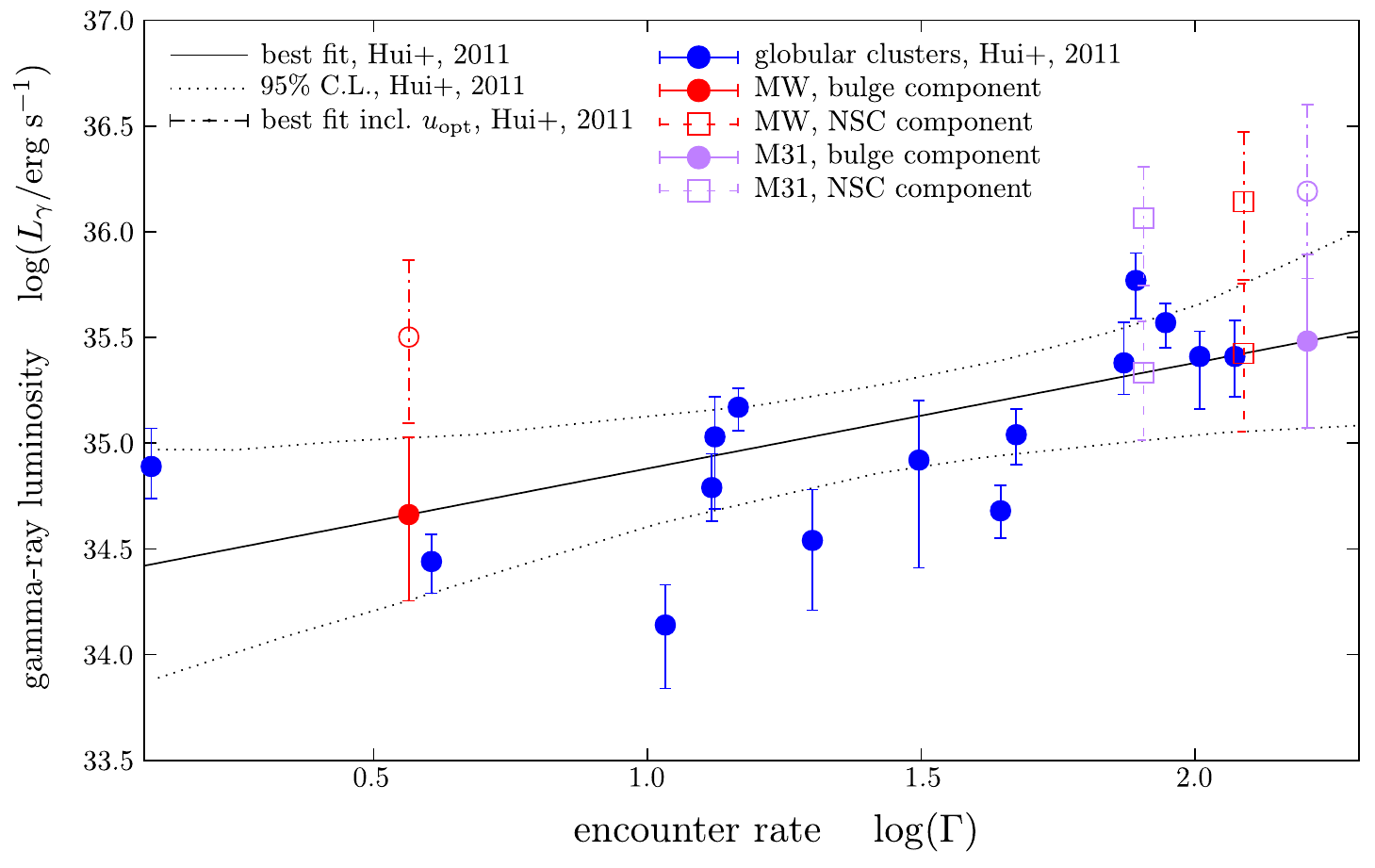}
\caption{Scaling relation of gamma-ray luminosity $L_{\gamma}$ of globular clusters with the stellar encounter rate, from (\cite{2011ApJ...726..100H}). The points for MW and M31 inner bulge regions and NSCs are added in this work, as described in Sec.~\ref{sec:globsclustscaling}. }\label{fig2}
\end{figure*}

\begin{table*}
\caption{}{Scaling relation  for the dynamical MSP component using globular cluster stellar encounter rates $\Gamma_e$. We show the best fit luminosity $L_{\gamma}\!\left(\Gamma_e\right)$ for each galactic component according to (\cite{2011ApJ...726..100H}) where we also provide the upper ($L_{\gamma,\textrm{max}}$) and lower ($L_{\gamma, \textrm{min}}$) limit on the luminosity with respect to the 95$\%$ C.L.~interval as displayed in Fig.~\ref{fig2}. Besides, we list here the luminosities ($L_{\gamma}\!\left(\Gamma_e,u_{\rm opt}\right)$) being derived by using the fit function in the same work that additionally includes the effect of an order-of-magnitude enhanced density of optical photons, $u_{\rm opt}$. } \label{tbl:luminosity}
\begin{center}
\begin{tabular}{l c c c c |}
\tableline\tableline
\\
System & $\log_{10} [L_{\gamma}]$ & $\log_{10} [L_{\gamma}]_{\rm{max}}$ & $\log_{10} [L_{\gamma}]_{\rm{min}}$ \\ 
\\
\tableline
\tableline
\\
 $L_{\gamma}\!\left(\Gamma_e\right)$ \\
 \\
\tableline
\tableline
MW, bulge $\vphantom{\intop_x^x}$ &  34.66 & 35.03 & 34.25 \\
MW, NSC &   35.4 & 35.8  & 35.1 \\
M31, bulge &  35.5 &  35.9 & 35.1 \\
M31, NSC & 35.333 & 35.6 & 35.01 \\
\tableline
\\
 $L_{\gamma}\!\left(\Gamma_e,u_{\rm opt}\right)$ \\
 \\
\tableline
\tableline
MW, bulge $\vphantom{\intop_x^x}$& 35.5 & 35.9 & 35.1 \\
MW, NSC &  36.1 & 36.5 & 35.8 \\
M31, bulge & 36.2 & 36.3 &  35.7 \\
M31, NSC &   36.1 & 36.6 & 35.8 \\
\tableline \tableline
\\
\\
\\
\end{tabular}
\end{center}
\end{table*}
\subsubsection{Globular cluster scaling relation} \label{sec:globsclustscaling}

In order to constrain the dynamical formation term from Eq. \ref{eq:primdyn}, we need to  determine the value of the normalization $\kappa$. We use the observed correlation of gamma-ray luminosity of globular clusters with the stellar encounter rate $\Gamma_e$ in these systems (\cite{2011ApJ...726..100H}):
\begin{align*}
\log{L_{\gamma}} = m \log{\Gamma_e} + c\textrm{,}
\end{align*}
with $m=0.50\pm0.16$ and $c = 34.38\pm0.25$. For this calculation, we also treat the NSC of each galaxy as an individual gamma-ray source besides the actual galactic bulge because of the enhanced stellar density therein which boosts the expected luminosity due to its $\rho_{\ast}^2$-dependence. 

Since the stellar encounter rate in a system of interest is proportional to the volume integral of $\rho_{\ast}^2/\sigma$ (\cite{Bahramian:2013ihw}), we use the stellar mass densities and velocity dispersion values as detailed in Sec.~\ref{sec:density} to derive an estimate of the rate. Except for the MW's NSC, our approach is the following: We normalize the resulting estimated rate to the one of the globular cluster Omega Centauri. For this cluster, we assume King's stellar mass density profile and structural parameters as listed in the globular cluster catalog (\cite{1996AJ....112.1487H}).

As the stellar density profile of the MW's NSC is not well known, we apply a different method: An object whose stellar density follows a King's profile exhibits a stellar encounter rate described by i) the central luminosity density $\rho_c$; ii) the core radius $r_c$; iii) the velocity dispersion $\sigma$, according to: $\Gamma_e \propto \rho_c^2 r_c^3\sigma^{-1}$ (\cite{2011ApJ...726..100H}). In what follows, we take the total extension of the MW NSC to be 10 pc and assume its stellar density to follow a King's profile. (\cite{Schodel:2014gna}) found that the total luminosity of the NSC within this region is $\left(4.1\pm0.4\right)\times10^7\;L_{\odot}$ which yields a luminosity density of $\rho_c=4.1\times10^4\;L_{\odot}/\textrm{pc}^3$ and in the case of a King's profile, the core radius is about $r_c=2$ pc. As shown in (\cite{2015MNRAS.447..948C}), the velocity dispersion inside the innermost 10 pc of the MW varies to a large extent with a mean value of about $\sigma = 150\;\textrm{km}\,\textrm{s}^{-1}$. Finally, we adopt the approach of (\cite{2011ApJ...726..100H}) to normalize the resulting stellar encounter rate to the one of the globular cluster M4 whose parameters are given in this reference.

We show the scaling relations in the cases discussed above in Fig.~\ref{fig2} and summarize the numerical values in Tab.~\ref{tbl:luminosity}.

\subsubsection{Luminosity prediction from dynamical formation} 

\begin{figure*}[!ht]
\centering
\includegraphics[width=.65\textwidth]{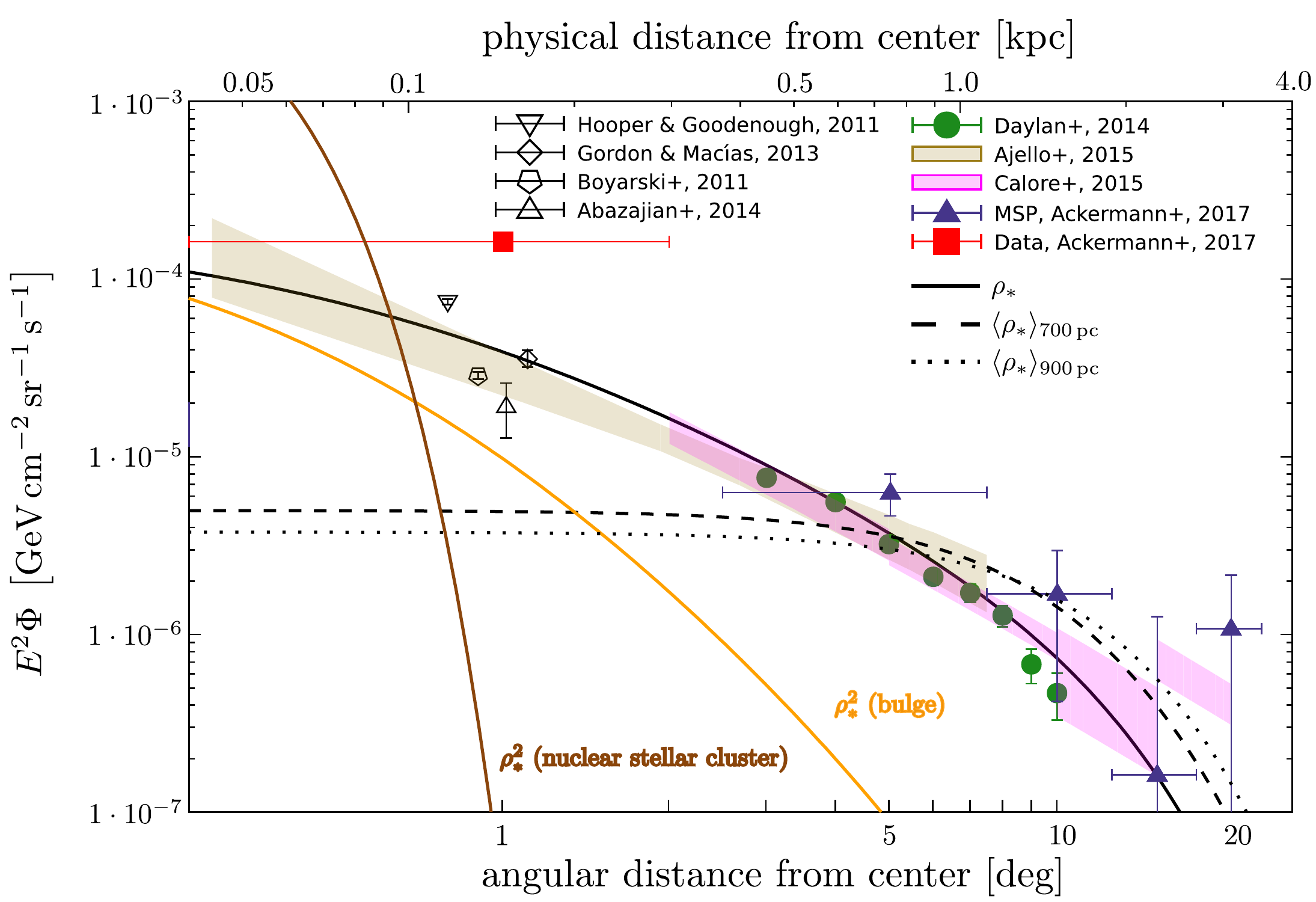}
\caption{Radial profile of the GCE emission as derived in a series of works, shown together with the predicted flux from a population of primordially formed MSPs in the Galactic bulge for three different MSP kick velocity smoothing kernels $d =$ 0 pc (solid), 700 pc (dashed) and 900 pc (dotted) applied to the stellar density in the MW bulge. The uncertainty on the normalization of the predictions is given in Eq. \ref{eq:Lmwprim}, with the actual curves displayed only reporting the best fit agreement to \cite{Calore:2014xka} within the allowed predicted range. We also plot our estimate for the maximal contributions of dynamically formed MSPs in the Galactic bulge (yellow) and NSC (brown), corresponding to the maximal allowed values for $L_{\gamma}\!\left(\Gamma_e\right)$ from Tab.~\ref{tbl:luminosity}. In both cases, no MSP kick velocity smoothing kernels have been applied (see discussion in Sec.~\ref{sec:refined_model}). For all curves, the LAT point-spread-function has also been accounted for via a  Gaussian kernel of width $1^{\circ}$, appropriate for an energy of about 2 GeV.  The PSF-induced morphological change is only dominant for the NSC contribution.}\label{fig3}
\end{figure*}

In our model, for both MW and M31 the maximal contribution from the dynamical formation ($L_{\gamma}\!\left(\Gamma_e\right)$) relative to the primordial gamma-ray luminosity is expected to be at the percent level (see Tab.~\ref{tbl:luminosity}). This does not necessarily mean, however, that these terms are irrelevant: 

In the case of the MW, it is possible that  the inner region GCE emission is significantly if not dominantly contributed to by this term, notably if the smoothing effect due to migration of the primordial MSP population turns out to be important. We illustrate this in 
Fig.~\ref{fig3}, representing the radial profile of the MW's GCE, predicted for primordial formation for the cases $d_1=$ 0 pc, 700 pc and $900$ pc together with the expected dynamical formation terms, due to the bulge and NSC stellar density. We show the {\it maximal} luminosity of such components allowed based on the globular cluster scaling relation with the stellar encounter rate given in Tab.~\ref{tbl:luminosity}. Note, however, that our rough approach to derive the stellar encounter rates is not sensitive to any small scale effects within the NSC (as resolved in $N$-body simulations, e.g.~\cite{2017arXiv170801627A}) which can lead to an enhanced gamma-ray emission. Thus, we might underestimate the maximally expected luminosity of galactic NSCs. Nonetheless, a much larger NSC contribution is already excluded by the red point in Fig.~\ref{fig3} that indicates the total GC emission within the innermost 2$^{\circ}$.

Even in the case of M31, some role cannot be dismissed, given the very large uncertainties affecting in particular M31 NSC contribution: A simple estimate of the stellar encounter rate, based upon the parametrization of the NSC according to \cite{2012A&A...546A...4T}, yields a luminosity which is comparable to what is expected from primordially formed MSPs in an object of the same mass as the NSC. 
An alternative view on the contribution of the dynamical formation to the total M31 luminosity is provided by the Chandra's observation of LMXBs in M31 (\cite{Voss:2007hj}). Using Chandra observations of M31 and simulations of LMXBs formation, \cite{Voss:2007hj} argue that out of 80 LMXBs observed in the inner region (inner 12 arcmin $\sim 2.7$ kpc) of the M31 bulge, about 20 ($\sim 25\%$) are believed to be dynamically formed. The main criteria used to identify the dynamically formed LMXBs was the observation that a fraction of these objects follows the $\rho_{\ast} ^2$ radial distribution. At the same time, in the inner 10 degrees of the MW $\lsi 20$ LMXBs were discovered in the INTEGRAL data (\cite{Haggard:2017lyq}), displaying no clear sign of $\rho_{\ast} ^2$ distribution, the main signature of a dynamical formation. This argument is not conclusive, in particular given the significantly different lifetimes of MSPs ($\sim 5$ Gyrs) and LMXBs ($\lsi 1$ Gyr, \cite{2013ApJ...764...41F}). A sharper quantitative comparison would require knowledge of the formation history of the systems. However, the evidence of a significant number of dynamically formed LMXBs in M31 might suggest that the dynamical formation is more important in the inner region of M31 than in the MW bulge.
Further  considerations of the connection between LMXBs and MSPs are reported in Sec.~\ref{sec:MW_bulge_disk_bias}.

\subsection{Corrections to the naive luminosity scalings} \label{sec:corrections-lum}

\subsubsection{Optical photon density induced MSP bias in bulges with respect to globular clusters}\label{uphot}
We did not account the effect of the optical photon density field. It has been suggested that the gamma-ray luminosity of globular cluster does correlate to some extent with the optical photon density $u_{\textrm{opt}}$  (\cite{2011ApJ...726..100H}):
\begin{align*}
\log{L_{\gamma}} = a_1 \log{\Gamma_e} + a_2 \log{u_{\textrm{opt}}} + a_3\textrm{,}
\end{align*}
with $a_1 = 0.42\pm0.17$, $a_2 = 0.62\pm0.29$  and $a_3 = 34.12\pm0.29$.  If this correlation reflects a physical process, such  as the availability of target photons for inverse-Compton, the correlation may be a symptom of a universal phenomenon: Until now, it might have been highlighted only in MSP globular cluster (as opposed to local MSP in the disk) because the background interstellar light is much higher in the (star and thus star-light rich) globular cluster environments. A similar correlation would enhance the emission of MSP from the bulge with respect to the disk. The effect is expected to be present both for primordially formed MSPs and  for dynamically formed MSPs. For the primordial objects, whose emission dominates in most of the bulge, one expects only a mild enhancement. Yet, this may be important in altering the shape of the emission profile, steepening it and thus counteracting to some extent the broadening effect of pulsar kick velocities. For dynamically formed objects, the impact on both the normalization of the signal should be more easily visible, the morphology being much harder to characterize given the Fermi-LAT PSF.
In general, the energy density of the soft photon field varies significantly in the central regions of MW and M31, suffering from large uncertainties. For illustration purposes, we show how large this effect could be if we were to adopt $u_{\textrm{opt}}$ ten times higher than in the Terzan 5 globular cluster, situated $\sim 2$ kpc from the center of the MW. The factor of 10 is a generous maximum error motivated by Fig.~2 in (\cite{Porter:2017vaa}). The results are listed in Tab.~\ref{tbl:luminosity} as $L_{\gamma}\!\left(\Gamma_e,u_{\rm opt}\right)$, suggesting an increase of about a factor seven, leading to a sizable (although unlikely dominant) dynamical term in the bulge.

\subsubsection{Gravitational  MSP bias in bulges with respect to globular clusters}
Another possible reason for a boosted dynamical term in bulges, compared with estimates based on globular clusters, is that bulges retain a larger fraction of the MSPs: this may turn to be the case if the intrinsic velocity dispersion of MSPs is large, but only relatively slow ones can be retained by the shallow gravitational potential of clusters.
 As the gravitational well in the central region of a galaxy is much deeper, the bulge might retain a higher fraction of its MSP. A relatively large dispersion may also ameliorate the agreement between a model expected to scale as $\rho_\ast^2$ with the apparent morphology of the GCE, seemingly scaling as $\rho_\ast$. Needless to say, such a conjecture would need to be tested with more elaborate models accounting for the phase-space distribution of the MSP. 
 
 \subsubsection{MSP bias in bulges with respect to disks} \label{sec:MW_bulge_disk_bias}
The determination of the local MSP luminosity function enters as a crucial ingredient in our estimate. Alternative derivations of the luminosity functions are broadly in agreement with what we used within errors (see for instance the results from~\cite{Ploeg:2017vai} which we compare with in  Fig.~\ref{fig4}), 
but the fiducial value can be even an order of magnitude lower  than what we adopted~(\cite{Bartels2017}).
The main reason for the broad range of these results boils down to the estimated local MSP detection incompleteness, which in turn depends on the chosen parametrization of the luminosity function, on the uncertainty on the distances to MSPs, their assumed spatial distribution templates, their gamma-ray spectrum, and the estimate of the (Galactic-direction dependent) Fermi-LAT sensitivity threshold. These are all ingredients for which often reasonable but certainly not identical choices have been adopted in the literature.

An intrinsic limitation of the scaling with stellar mass adopted here is due to the fact that the derivation of the luminosity function does not exactly predict a constant luminosity over stellar mass ratio within the MW disk, to start with. However, the agreement of both the GCE energetics and morphology with our simple scaling is tantalizing. Notice that in \cite{Bartels2017}, where the local MSP luminosity function is not used to predict the GCE, the scaling is found to be valid within different bulge sub-components. 
Should the local MSP luminosity function be revised downwards, or should an improved sensitivity to M31 disk emission turn into a null result, the situation may thus appear puzzling. A possible solution to such a conundrum may be found into a more complex scaling of the emission than just with the stellar mass. One may expect, indeed, an impact of stellar ages on the MSP luminosity, via a number of effects: For instance, the value of $M_{*}^{\rm MW}$ used in our prediction refers to the total stellar mass including old and young populations of stars. One may argue that MSPs are more likely to be found in old stellar populations, as the majority of MSPs has lifetimes of about $\sim5$ Gyr. In the MW disk, however, there are regions of active star formation. According to the Gaia-ESO survey, the fraction of stars in the disk younger than 5 Gyr is about 40\% (\cite{2017AA...603A...2M}). The actual bulge ratio $L_{\rm b}/M_{\rm b}$ is thus expected to be about 1.7 times larger than our naive estimate based on the disk luminosity function, since the bulge is largely dominated by old stars. Note that such a bias would still preserve the scaling with density within the bulge, which seems observationally established. On the other hand, the estimates in Sec.~\ref{sec:ModelMW} and Sec.~\ref{sec:M31-toy-model} assume that the MSP luminosity function is independent from the age of the stellar population. Detailed estimates are lacking, but one may anticipate appreciable corrections to the above naive scaling.

Complementary to our assumption of rescaling the MW's MSP luminosity function based on the stellar mass of the object under study, we could have used the number of LMXBs therein, since LMXB  are thought to be progenitors of MSPs. Such an approach has been followed in~\cite{Haggard:2017lyq}, where the authors analyzed the available collection of observed LMXBs in the MW and its globular clusters to perform such a rescaling. They report that about $10^3$ LMXBs in the bulge region of the MW would be needed to fully account for the GCE while only $< 100$ of them were found. They estimate that $\lesssim 23\%$ of the GCE can be attributed to an unresolved MSP population. In our opinion, even an apparent discrepancy of less than one order of magnitude, does not rule out our model for a number of reasons.  
First of all, the characteristic lifetimes of LMXBs and MSPs differ by more than one order of magnitude, which prevents a detailed comparison without knowing the time-dependence of the two formation histories: LMXBs associated to the bulge population of MSP have already faded away.
Additionally, it is questionable if a {\it perfect} scaling between LMXBs and MSP should be assumed, since alternative formation channels have been proposed for both systems. For example, as an important caveat to a simple scaling, we mention here that there are different types of LMXB, notably hydrogen-rich donor LMXB and  hydrogen-poor donor LMXB, known as  ultracompact  X-ray
binaries (UCXB).  In the bulge-specific study~\cite{vanHaaften:2015wta} it has been argued that, although the former are much easier to observe given their luminosity and duty cycle, UCXB have in fact a higher formation rate, and would be the preferential progenitors of MSP, while being much fainter and difficult to identify. 

An interplay with the role of the background photon field (Sec.~\ref{uphot}) is also possible: if this ingredient also affects the MSP emission in the primordial populations in the disk and bulge, it would certainly boost the expected bulge emission with respect to the
na\"{i}ve predictions based on the disk sample, given the higher background $u_{\textrm{opt}}$ in the bulge than in the disk, as reported in \cite{Porter:2017vaa} for the MW and in \cite{Groves:2012cu} for M31.

\begin{figure*}[!ht]
\centering
\includegraphics[width=.75\textwidth]{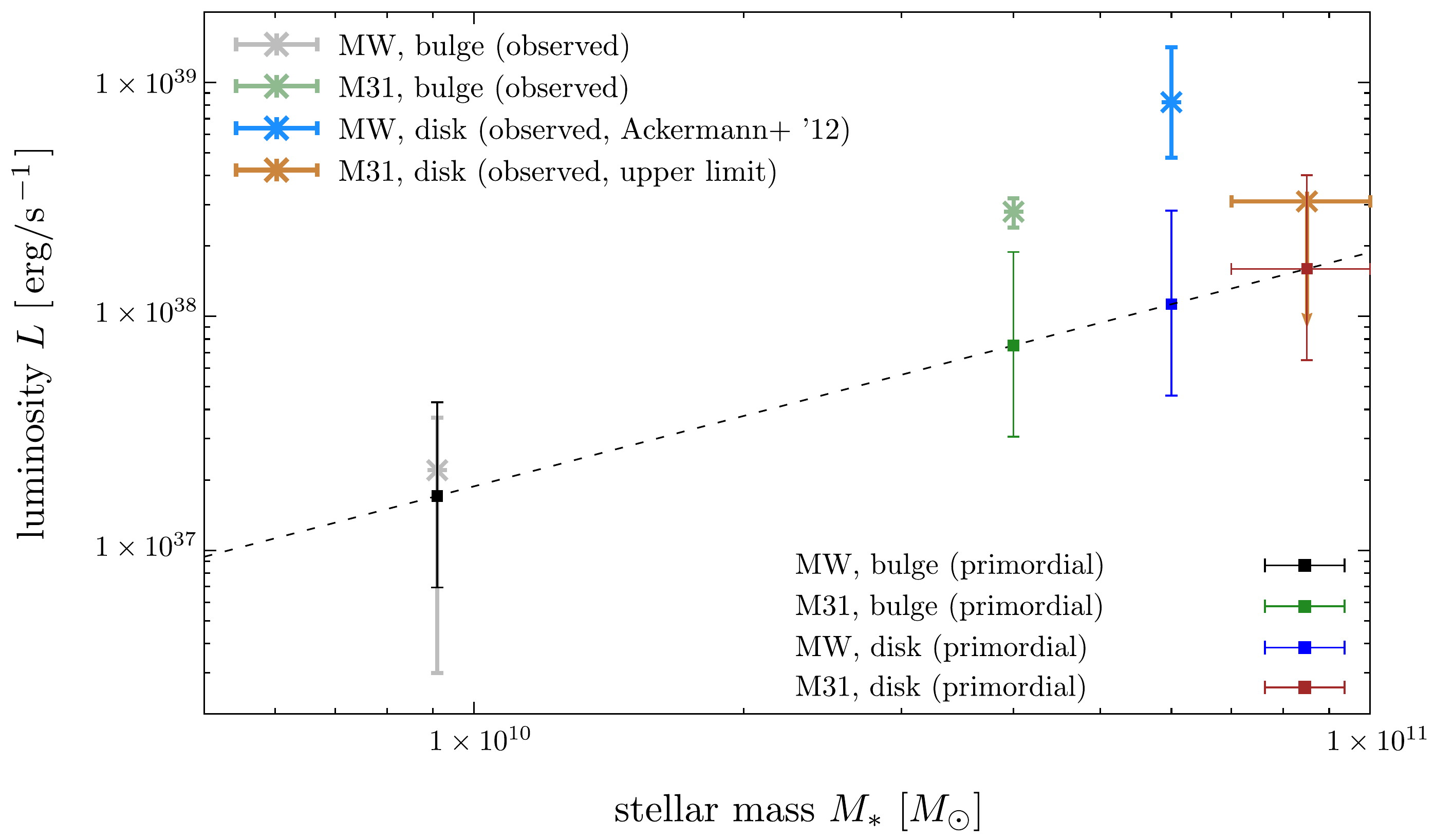}
\caption{Summary plot of the predictions of our toy model in comparison with the respective observed luminosities as reported in the text. We show as a black, dashed line the luminosity prediction for primordially formed MSPs in a region of particular stellar mass $M_{\ast}$ as derived from the local MSP luminosity function (Eq.~\ref{eq:prim}). 
}\label{fig4}
\vspace{0.65cm}
\end{figure*}

 \subsection{Alternative origins for the bulge MSPs}
It may be also that alternative mechanisms for populating the bulge with MSP are at play. For instance, it  has been proposed that a population of MSPs in the 
inner Galaxy was deposited from disrupted globular clusters, that were brought down through dynamical frictions process and tidally stripped (\cite{Brandt:2015ula}). 
Simulations of tidal disruption of globular clusters (\cite{Gnedin:2013cda, Arca-Sedda:2017qcq, Fragione:2017rsp}) demonstrate that the NSC could effectively form because of this deposited mass. The recent study~\cite{2017arXiv170903119A} seems to confirm the viability of a scenario where the
X-ray and the $\gamma$-ray high-energy emissions  from the innermost Galaxy are related to the NSC formation via tidal disruption of globular clusters,
and the associated deposition of MSPs and  Cataclysmic Variables. However, at larger distances (100 pc to 1 kpc), the disrupted stellar mass distribution tracks the  globular cluster distribution set as the initial condition for the simulation (\cite{Gnedin:2013cda}), so the model is not directly predictive, as far morphology is concerned.  

The recent work \cite{Fragione:2017rsp} suggests however that both the intensity of the GCE signal and the normalization and distribution of surviving globular clusters in the inner Galaxy support this scenario, addressing some objections to  \cite{Brandt:2015ula} raised in \cite{Hooper:2016rap}. The authors compare the MSP luminosity function of globular clusters with the findings from MSPs in the MW disk. They conclude that MSPs from old stellar populations like those found in globular clusters can only account for a few percent of the observed GC emission.  If this scenario is correct, then the rough agreement of our disk luminosity scaling with the GCE would appear to be a coincidence. Also, at the moment no prediction for the M31 bulge emission is available, lacking dedicated simulations. The model does not address the disk luminosity scaling either, of course (hence no prediction for M31 disk emission).

\section{Summary}\label{sec:summary}
Since the start of the Fermi-LAT mission, more than a hundred gamma-ray MSPs were discovered, proving that they are ubiquitous in our Galaxy. Over time, evidence is also accumulating that these objects are responsible for the diffuse emission in the inner region of our Galaxy (the GCE). More recently, an extended emission has also been observed from the direction of M31, showing broad features similar to the GCE. In this work, focusing on in situ formation of MSPs, we used the properties of known local MSPs to describe the primordial MSP formation. This simple, parameter free scaling predicts then the MSP luminosity in the bulges of the MW and M31. 
Despite its simplicity, this model fits remarkably well both the energetics and the morphology of the observed GCE, as summarized in Fig.~\ref{fig4}. 
Regarding the model's predictions for M31, the estimated luminosity due to primordial MSPs (Eq.~\ref{eq:M31primlum}) is typically expected to contribute only about a quarter of the detected emission, although a dominant contribution cannot be excluded at the 2$\sigma$ level, given the large uncertainties. A prediction of the model is that the M31 disk emission due to MSP is not far below the present upper bound, a fact that might allow for an observational test in the not so distant future. 

We have also discussed some refinements of the simple model above. In particular, we used a scaling based on globular cluster gamma-ray luminosity with the stellar encounter rate to describe the dynamical MSP formation in the bulge and the nuclear star cluster. This component is expected to contribute to the bulge emission only at a level $\lesssim 5\%$ (cf.~Fig.~\ref{fig2}), but---depending on the size of the smoothing effect due to the MSP kick velocity distribution---may be of some importance in explaining the signal morphology in the inner region of the MW. We also commented on some other possible effects which may lead to violations of the simple scaling used and thus different expectations, and briefly discussed alternative models for populating the bulge with MSP, in particular the scenario of globular clusters  disruption at an early stage of the MW evolution.

In the near future, an improvement in the local MSP luminosity function determination, together with improved observational studies of both the non-thermal activity of sources in the MW and M31 bulge and M31 disk emissions may narrow down the current spectrum of possibilities. In our opinion, however, major advances will require a sizable reduction of the uncertainties on the MSP formation mechanisms: In primordial scenarios, the capability of newborn pulsars to stay tied to the original binary system despite the kick velocity at birth needs to be better understood. In dynamical formation scenarios, a quantitative understanding of the capture of either rogue pulsars by binaries and of the probability of pulsars in binary systems ending up in MSP via stellar encounters is also crucial. The variability of these processes with the environmental properties (e.g. in Globular Clusters vs disks vs bulges) is another open question, like is the survival or formation of MSP in Globular Cluster disruption events. If coupled to non-gravitational gas and stellar dynamics, these studies would also be beneficial to clarifying issues like the link of the MSP phenomenon to the LMXB one, for instance. Eventually, we are confident that the synergy of such studies will shed further light on these fascinating gamma-ray observations emerged over the past few years.

\vspace{0.5cm}

\section*{Acknowledgements}

We thank F. Calore, C. Weniger, and especially R. Bartels, A.~Clerici and D. Malyshev for discussions and feedback on a preliminary version of this manuscript.

MDM acknowledges support by the NASA {\it Fermi} Guest Investigator Program 2014 through the {\it Fermi} multi-year Large Program N. 81303 (P.I. E.~Charles).
X.H. is supported by the National Natural Science Foundation of China through grant 11503078.
PM acknowledges support from the Centre National d’Etudes Spatiales.
P.S. acknowledges support from Agence Nationale de la Recherche under the contract ANR-15-IDEX-02, project "Unveiling the Galactic centre mistery", GCEM (PI: F. Calore).

\appendix
\subsection{Spectrum of the M31 emission} \label{sec:spectandprofile}

In PaperI it was found that the spectrum of the M31 emission can be fit reasonably well with a power-law (PL) spectrum and a power-law with an exponential cut-off (PLEC) $E^{-\Gamma} \text{exp} [-E/E_{\text{cut}}]$. The best fit parameters of the PLEC fit were found to be $\Gamma= 2.1\pm 0.2$ and $E_{\text{cut}}=5.3\pm4.9$. This is broadly consistent with the spectrum of MSPs which is also found to be in the PLEC form, with $\Gamma=1.57^{+0.01} _{-0.02}$ and $E_{\text{cut}}\sim 3.78 ^{+0.15} _{-0.08}$ GeV, (\cite{Cholis:2014noa}).

In addition to the 'prompt' gamma-ray emission, it is well accepted that young pulsars are an efficient source of electron-positron pairs (called just 'electrons' in what follows) and they are considered as one of the main suspects in explaining the high energy positron emission as measured by the PAMELA and AMS-02 satellites (e.g. (\cite{Hooper:2008kg})). \cite{Venter:2014ata,Venter:2015gga}) investigate the MSP population as a source of Galactic electrons. While the electron injection spectrum is largely unconstrained, it has been argued that the efficiency of gamma-ray emission of MSPs, i.e. ratio of gamma-ray luminosity to total spin down rate,  is near $10\%-20\%$ (\cite{Johnson:2012rk}) which leaves a possibility that a comparable portion of spin-down energy is taken up by electrons. 

Electrons with energy in the 10 GeV range lose most of their energy via the Inverse Compton (IC) scattering on CMB and interstellar radiation, yielding up-scattered gamma-ray photons, and synchrotron radio emission. In \cite{Petrovic:2014xra}, the contribution of such IC emission was estimated under the following two assumptions: i) the injected electron spectra follow a PLEC shape (cf.~\cite{Venter:2014ata,Venter:2015gga}) and ii) we fix the relative normalization of the photon and total electron fluxes by requiring an {\it equal} amount of energy distributed between the two species, see (\cite{Petrovic:2014xra}) for details.

In \cite{Petrovic:2014xra}, some of us modeled both the prompt and IC emission from MSPs, using the GALPROP code and found that the resulting spectra can well explain the emission at the center of our Galaxy -- both the 3 GeV 'bump' and its high energy tail (confirmed also by (\cite{Linden:2016rcf,OLeary:2016cwz})).  
 
Here we repeat this argument for the bulge of M31 assuming the strength of the magnetic field to be $20\upmu$G and a two times stronger interstellar radiation field than in the MW bulge. In Fig.~\ref{fig5} we show energy spectra averaged in the 5 kpc region of M31, for the benchmark values of the electron injection parameters from \cite{Petrovic:2014xra}. We see that while the M31 measurement is in agreement with the prompt MSP spectrum alone, a high energy tail of this emission is well motivated and could potentially be confirmed by next-generation gamma-ray probes such as the CTA (\cite{Acharya:2017ttl}). 

\begin{figure*}
\centering
\includegraphics[width=.65\textwidth]{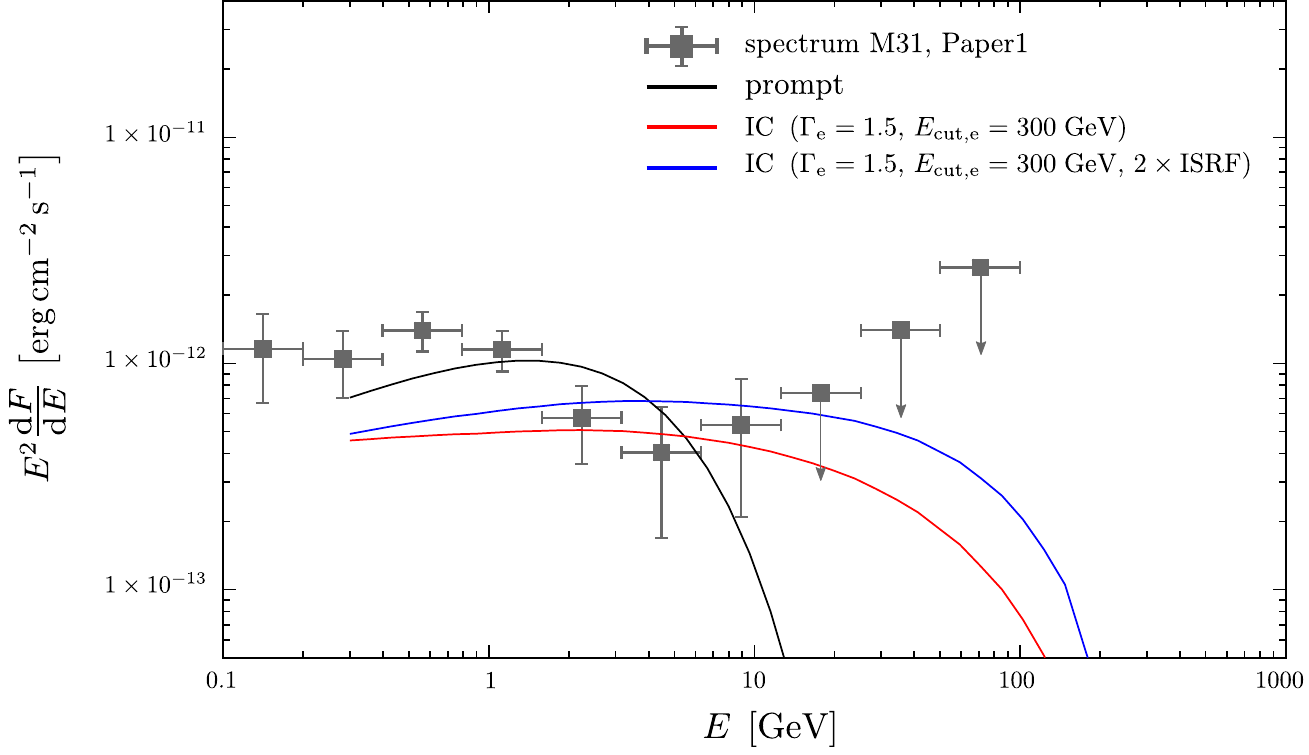}
\caption{The gray squares show the spectrum of M31 as reported in PaperI. The relative normalization between prompt and IC emission follows the scheme in (\cite{Petrovic:2014xra}) whereas the contribution from prompt emission is normalized to the data.\label{fig5}}
\end{figure*}

\bibliography{MSP_M31BIB}

\end{document}